# Cross-infrastructure holistic experiment design for cyber-physical energy system validation

Van Hoa NGUYEN[1], Quoc Tuan TRAN[2], Yvon BESANGER[1], Tung Lam NGUYEN[1,2],
Tran The HOANG[1], Cédric BOUDINNET[1], Antoine LABONNE[1], Thierry BRACONNIER[1], Hervé BUTTIN[2]
[1] Univ. Grenoble Alpes, CNRS, Grenoble INP, G2Elab, F-38000 Grenoble, France
[2] CEA-INES, 50 Avenue du Lac Léman, 73370 Le Bourget-du-lac, France
Email : van-hoa.nguyen@grenoble-inp.fr

*Abstract*— Strong digitalization and shifting from unidirectional to bidirectional topology have transformed the electrical grid into a cyber-physical energy system, i.e. smart grid, with strong interdependency among various domains. It is mandatory to develop a comprehensive and holistic validation approach for such large scale system. However, a single research infrastructure may not have sufficient expertise and equipment for such test, without huge or eventually unfeasible investment. In this paper, we propose another adequate approach: connecting existing and established infrastructures with complementary specialization and facilities into a cross-infrastructure holistic experiment. The proposition enables testing of CPES assessment research in near real-world scenario without significant investment while efficiently exploiting the existing infrastructures. Hybrid cloud based architecture is considered as the support for such setup and the design of cross-infrastructure experiment is also covered.

## I. Introduction

The European Energy roadmap [1] require higher penetration of distributed renewable energy resources (DRES) and more advanced energy management framework, in order to achieve the desired decarbonized scenario. The increasing number of DRES at distribution level (e.g. wind, photovoltaic, combined heat and power, etc.), however, may cause some bad influences to the performance of the power system (e.g. fluctuating voltage, unstable frequencies, reverse load flows, etc.) as well as may add various uncertainties (e.g. volatile DER generation) that cannot be neglected while operating the network. The traditional unidirectional power grid has to adapt to the bi-directional power flow due to the DRES generation being connected to the distribution level. Moreover, there is a worldwide tendency to develop micro-grid for resilient energy security. This, with the liberalization of energy markets and the regulatory changes, present the necessity of forming decentralized energy trading frameworks, which requires more adapted infrastructures as well as non-conventional technological support [2].

That topological change poses a great challenge to maintaining a good functionality for the power system and urges researchers to introduce more innovative technological solutions. On the other hand, with strong digitalization of the power system and advanced metering system shifting from experimental phase to deployment, the power grid has turned into a multi-layered cyber-physical energy system with strong interrelation and interdependency among different domains. It is necessary to restructure and to adapt the design and operation of the smart grid to such fundamental changes. The power grid, however, is a system with high complexity and strict security requirement. New technological propositions have to be rigorously validated before they can be deployed to the grid. It is mandatory to have a comprehensive and holistic validation to guarantee a reliable functionality of such cyber-physical energy system (CPES). The traditional methods of testing single component or domain is no longer sufficient for validation of such system of systems. The complex interaction among the domains needs to be taken into account[3]. The holistic validation approach also has to be rapid and automatable because more and more use-cases and applications of new technologies are being integrated to the grid.

Since a holistic validation methodology for smart grid considers the CPES as a whole and takes into account the interdependency of the domains, it requires a suitable complexity of validation environment (i.e. communication network, advanced metering infrastructure, various types of DRES). Due to the dimensions and the costs of realistic testing environment, one single local validation in only one research infrastructure is often not feasible, due to lack of expertise or equipment. An adequate approach is connecting already existing and established laboratory infrastructures with complementary focus of specialization and facilities [4]. Such a setup would provide the possibility to reflect the real CPES as close as possible, to efficiently exploit the research infrastructures and to rapidly transfer new developments. In order to achieve such coupling, interoperability among the infrastructures is identified as a fundamental requirement. In [5], a hybrid-cloud based supervisory, control and data acquisition (SCADA) system with platform-as-a-service

(PaaS) model of service deliverance has been proposed to be a suitable support for interoperability among research infrastructures.

In this paper, we envisage the approach of using holistic cross-infrastructure experiments for CPES assessment. The hybrid-cloud based support is considered and the design procedure of a cross-infrastructure holistic experiment is presented. The paper is organized as following: in section II, we consider the utilization of hybrid-cloud based support for holistic cross-infrastructure experiment. The interoperability model is then analyzed and we examine why such a support is suitable for CPES assessment in a cross-infrastructure context. In section III, possible cross-infrastructure validation techniques for CPES are introduced and the design of such experiments is discussed. A layered conceptual view of the cross-platform interface, along with the associated issues and recommendation, is presented. Finally, an exemplary case-study of interoperability of platforms PREDIS and PRISMES is presented in section IV, before the paper is concluded in section V.

## II. Hybrid-cloud based support for holistic cross-infrastructure experiment

### A. Coupling laboratory infrastructure for holistic experiment of CPES

Due to the increasing complexity of the CPES, an integrated approach for analysis and evaluation is necessary to support the forthcoming large-scale roll out of new technological developments. The holistic validation procedure needs to consider the CPES not only at component – single domain level, but also at system – multi-domain level. Such an approach does not exist up to date and is considered hard to achieve because of various open issues. Most importantly, the CPES is a system of systems including interdependent domains (power, ICT, heat, gas, etc.); it is necessary: 1/ to possess a combined expertise of different domains, which is not always the case of current laboratory system; and 2/ to develop corresponding research infrastructure with proper methods and tools, which implies a huge investment. While it is possible to use simulation in early phases of validation of different setups and configurations for CPES, in some case (e.g. simulated model of the hardware under test is not available), it is necessary to involve hardware prototypes and proper laboratory infrastructure. It is even of more importance to represent a near real-world environment in testing of such a critical element as the power system.

As previously mentioned a holistic experiment for CPES assessment requires a large scale validation scenario and cannot be done in one research infrastructure. Building a complete infrastructure with all necessary components and technologies is theoretically possible, but is realistically not a reasonable solution, in term of operation (not only the equipment, but also domain specific competences are needed), finance and organization. Besides, even among the laboratories of the same domain (e.g. electrical engineering), there are still difference in the particular research focus (e.g. power grid, renewable energy generation, control and command, etc.), thus different focus of infrastructure investment. It is therefore reasonable to establish an infrastructure coupling framework to efficiently use the existing equipment and combine them with the complementary counterpart from others to validate researches in a holistic and near real-world environment.

On the other hand, development of such a holistic validation framework for CPES would also benefit researchers in term of facilitating the replication of experiment and the verification of the validity of the results. As the experiment is cross-infrastructure, it is expected to apply or develop harmonized (and possibly standardized) experiment design and evaluation procedures [6].

### B. Requirement on interoperability of infrastructures

To enable the holistic cross-infrastructure experiment, it is required to achieve interoperability among the participating infrastructures. Interoperability is still a new subject in the domain of smart grid and several difficulties towards interoperability among research infrastructures can be identified: lack of a suitable interoperability model, possible differences in RI policies, lack of flexible information model covering both power and ICT domains, and lack of effort to harmonize the excessive numbers of communication protocols. Based on the SGAM model [7], we can identify five layers of interoperability among a consortium of research infrastructure (Figure 1).

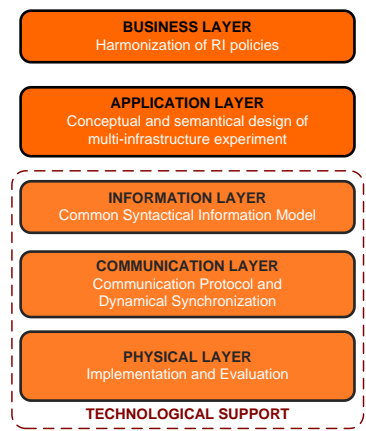

Figure 1: Five layers of interoperability among research infrastructures.

From the top layer, a harmonization of communication and sharing policies inside the consortium of participating infrastructures as well as agreement on intellectual properties should be done. In general, it establishes the legal support to implement the cross-infrastructure experiment. The applications layer deals with the conceptual and semantic design of multi-infrastructure experiment, based on the information of RI's capabilities. From a more technical point of view, interoperability requires the consortium infrastructures to share a common information model or at least to possess a conversion interface to the selected information model (e.g. CIM). The communication layer represents the harmonization of communication protocols as well as the aspect of synchronization, handling causality and latency compensation. It ensures a seamless communication among infrastructures, the good emission and reception, and the correct interpretation of received information. Finally, the physical layer represents the actual interoperable implementation as support for the other layers, including also the aspect of evaluation of performance.

In general, each research infrastructure has already its own configuration in term of equipment, communication and protocol set up. For cross-infrastructure holistic test, it is therefore judicious to aim to interoperability at information layers and above, to save effort on adaptation of local existing infrastructures.

### C. Hybrid-cloud based support

In order to enable interoperability among research infrastructures and to establish a seamless access to shared resources, it is necessary to integrate the SCADA architecture of the participant's infrastructures. The integrated infrastructures are most of the time distanced from each other, so latency and information loss in communication are potential issues to the performance of the joint platform, thus, new architectural approach with ICT integration is required. In [5], a hybrid cloud based SCADA architecture was proposed to address the problems of interoperability and holistic cross-infrastructure experiment among research infrastructures (Figure 2).

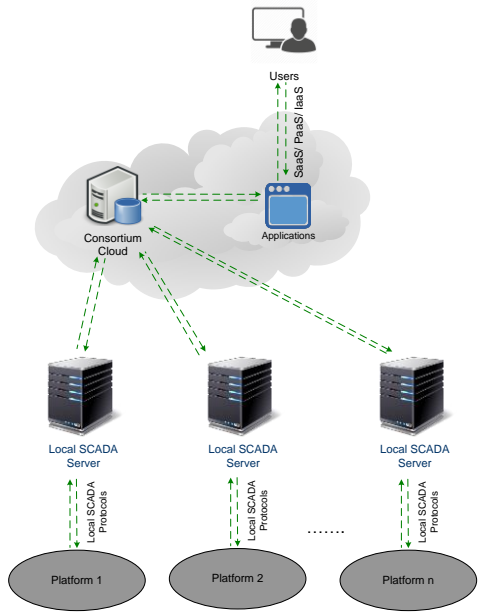

Figure 2: Hybrid cloud for interoperability of Research Infrastructures

The architecture is based on the concept of hybrid cloud (in the sense that only some applications are available via the consortium cloud server) and platform-as-a-service (PaaS) model of service deliverance (i.e. user do not have full control over the infrastructure, but only have access to a selection of commands). The local SCADA servers are in charge of supervision via the corresponding intranet. The critical SCADA tasks are solely located on-site and are controlled by the local SCADA servers, PLC or RTU. It is imperative because of the requirements of high responsiveness and low latency for such applications. On the other hand, the databases of the local SCADA server is replicated into a common standardized database located on the consortium cloud. Model translation and mapping of variables may be required if the local SCADA servers do not share a same information model. Latency tolerant applications such as AMI, DMS, VAR optimization and outage management can also be deployed over this server. Using PaaS delivery model, it is possible to remotely demand to launch a certain function (setting values, starting simulating, etc.) and visualize the result, provided the authorization of the platform operators. This property enables the possibility to launch experiment on the shared resources, without having to come to the platform in person or eventually, combine remotely located resources in a holistic experiment.

The architecture enables the possibility to cooperate assets and expertise from various RI in multi-platform experiments for CPES assessment.

### III. DESIGN OF CROSS-INFRASTRUCTURE EXPERIMENT

We discussed in precedent sections the necessity to establish a holistic framework for CPES assessment and the possibility to develop cross-infrastructure experiment over a support of hybrid-cloud architecture. We present in this section the potential cross-infrastructure experiments that can be deploy on such architecture. The design of cross-infrastructure interface for such experiments is also examined and the layered issues are identified.

### A. Cross-infrastructure co-simulation and hardware-in-the-loop

Co-simulation and remote hardware-in-the-loop (HIL) are the two types of experimental technologies for CPES assessment that can be considered for deployment over the cross-infrastructure hybrid-cloud architecture. While co-simulation can be used as a combination of expertise, remote HIL experiment can be used in case of sharing experimental equipment or combination of equipment from various infrastructures in a holistic testing framework.

Co-simulation in the field of CPES assessment is mainly used to investigate the interconnection of power system and communication network simulation [8]. It is essential due to the lack of a computing tool that can handle both continuous model (power system) and discrete event model (communication). Co-simulation provides also the possibility to investigate cyber-security impact to power system (i.e. denial-of-service protection, confidentiality and integrity testing).

Hardware-in-the-loop technique is widely used in CPES assessment framework, especially for testing of DER devices.

In this approach, a real hardware is coupled with a simulation tool in real-time. This setup allows experimenting of hardware/software components under realistic and scalable conditions (e.g. faulty and extreme conditions) without having to connect them to the real power system, which is not always a feasible option due to security issue. Besides, the technique allows replacing the error-prone or incomplete models with their real counterparts and therefore improves the accuracy of the experiment. Depending on whether signal or power flow is exchanged between the hardware and the real-time simulator, HIL can be classified into control-HIL or Power-HIL. In the case of PHIL, a power interface is necessary (Figure 3).

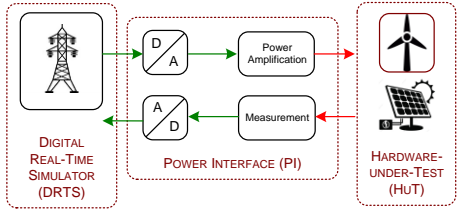

Figure 3: General architecture of a PHIL experiment.

Providing new possibility for CPES testing, the two techniques are also considered to be bought together in a holistic framework for CPES assessment [9]. Combining the two technologies allows an in-depth study of CPES with realistic behaviors from hardware equipment under a variety of complex environment, co-simulated by several simulators from different domain. Moreover, it helps to improve the accuracy of the experiment as different models are connected to their appropriate solvers (i.e. power system to continuous and ICT to discrete event solver; steady state solvers for large scale simulation and real-time capable solvers for transient simulation).

This integration however faces several challenges: the synchronization of continuous and discrete event models of computation, concurrency of data flow and real-time/non real-time harmonization. As for the deployment of a combination of co-simulation and HIL in a cross-infrastructure experiment, it is expected to address the co-simulation interface. The distance among the infrastructures will introduce latency to the communication. It may lead to problem in synchronization and for HIL experiment, may violate the real-time constraint. In order to maintain the accuracy of the test, it is suggested to keep the HIL interface intra-platform.

The hybrid cloud based SCADA architecture can be used as a support for integrating hardware/software from various infrastructures, as well as a base for deploying master algorithm. In fact, the hybrid cloud server can act as a common buffer for information exchange among participating entities, whether they are simulation results from software or realistic measurement from the SCADA system. The users get access to the data and are capable of managing the master algorithm via PaaS or SaaS model.

This approach can be considered as an asynchronous integration of HIL to co-simulation. In fact, the HIL interface is located intra-platform while best effort interconnection is established between the simulators and the cloud server. From then, the user has the capability to configure the test accordingly by using real-time connections or using a formalism wrapping technique such as waveform relaxation method [10].

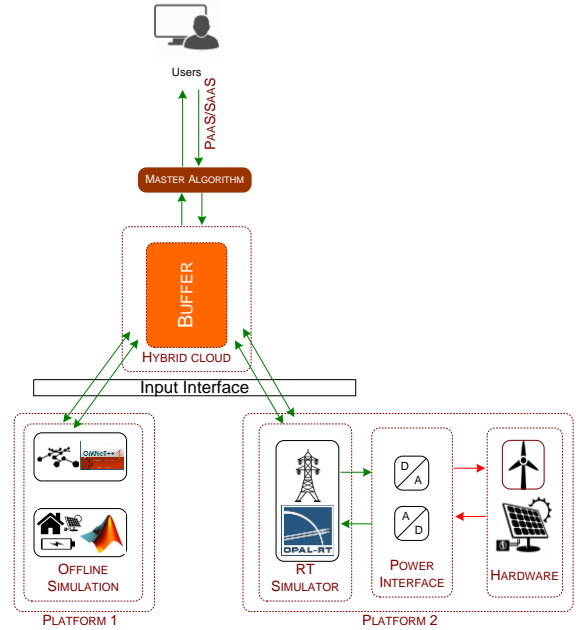

Figure 4: Cross-Infrastructure Experiment using Hybrid cloud-based server

Another option is using a common message "hub" to govern exchanged signals and to route the messages. The synchronization aspect can be done via conservative approach (as in OPSIM[1]) or neglected (as in LabLink architecture [11] - Figure 5). The semantic communication via master orchestrator requires however manual configuration.

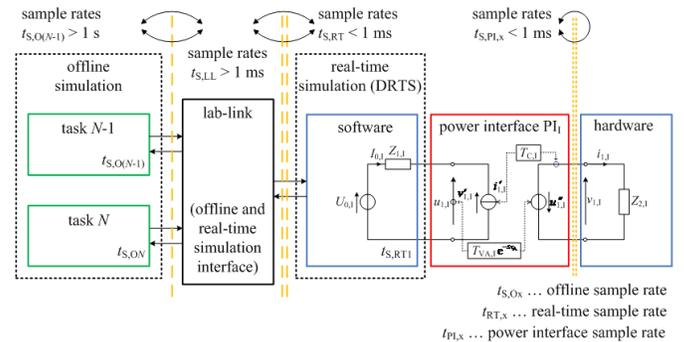

Figure 5: Lablink Architecture with recommended sample rates [11]

The above architectures provide the possibility to combine co-simulation and HIL experiments in a cross-infrastructure manner and to combine different models to their appropriate solvers in a large scale holistic experiment. In order to enable such experiment, it is required to have a machine readable and semantically coherent information model for the infrastructures.

---

[1] www.opsim.net – developed by Fraunhofer IWES

### B. Implementation of machine readable information models

Since the holistic experiment collects information from various sources with potentially different formats, in order to guarantee the interoperability among infrastructure as well as ensure seamless performance of the experiment, we need to achieve interoperability among the participants; as aforementioned, in information layer and above. It also helps to facilitate planning and to conduct analysis in case of a fault. In general, it is recommended to adopt a machine readable information model and experimental design. A machine readable model of the relevant experimental aspects is important to enable automated orchestration and configuration of experiment.

It is necessary to emphasize the two different semantical models here: the first one is the electrical domain information model, representing the involved equipment in the system configuration and their interconnection and secondly, the semantical model representing the interaction among different test-element as well as the causality and transition among test-stages. The latter one can be considered of higher layer of abstraction. Therefore, it is not sufficient to employ only a single information model for the test, but a combination of two or more information model, in order to fully assure the semantical coherence. It is suggested in [12] that the CIM (IEC 61970 and IEC 61968) is a suitable common information model for interoperability among RI in electrical domains and can be used to configure the consortium database. An adaptive method to deploy CIM over the consortium cloud, given the CIM/XML//RDF description of the participating RI, was proposed in [13]. This innovative way ensures interoperability between partner platforms and provides support for a holistic multiplatform approach to smart grid evaluation. As for experiment modeling, the testing description standard TTCN-3 can be considered [10]. The use-case approach and the 62559 template from SGAM model[7] is also a viable option.

Finally, it is important to correctly and systematically set up the cross-infrastructure interface. In fact, in the proposed architecture, each individual institution can maintain their own communication infrastructure and interoperability is actualized up from information layer, via mean of the cross-infrastructure interfaces. Based on the structuration proposed in [10], we highlight the layers and notable associated issues in setting up a cross-infrastructure interface for a holistic experiment:

- **Conceptual layer:** Generic structure of the framework and meta-modeling of the components (e.g. conceptual configuration of the experiment).
- **Semantical layer:** Signification of the individual models and their interaction (may be represented by test description standards e.g. TTCN3)
- **Syntactic layer:** Formalization of individual models in the domains, harmonization of the difference among the models (may be done via using domain specific information models: e.g. CIM, IEC 61850).
- **Dynamic layer:** Order of execution of processes, synchronization and causality of model (e.g. conservative or optimistic approach), harmonization of different models of computations (continuous vs. discrete event).
- **Technical layer:** specific implementation of the interface, protocols, evaluation of performance.

## IV. CASE-STUDY: HOLISTIC CROSS-INFRASTRUCTURE EXPERIMENT BETWEEN PREDIS AND PRISMES PLATFORMS

A case-study of interoperating two smart grid platforms in a cross-infrastructure framework, from the ongoing French project PPInterop II, is introduced. PREDIS (Grenoble INP) is a smart grid experimental platform consisting of 11 smaller technological platforms covering various aspects of smart grid, notably the re-scaled distribution grid, the supervision center, smart building and the real-time platform. PRISMES (CEA INES) consists of a low voltage micro-grid and a wide variety of DRES, controllable loads, electrical vehicles and storage systems. Coupling these two complementary platforms allows setting up holistic experiment of the CPES in various scenarios (high penetration of DRES, impact of EV to power system management, etc.). The two platforms are located 70km apart (Grenoble - Chambéry) and PPinterop II aims to make these platforms interoperable, to remotely monitor and control in real time the available tools, as well as to conduct cross-infrastructure experiments.

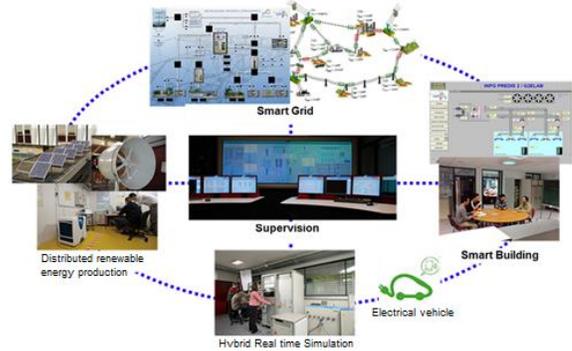

Figure 6: PREDIS platform

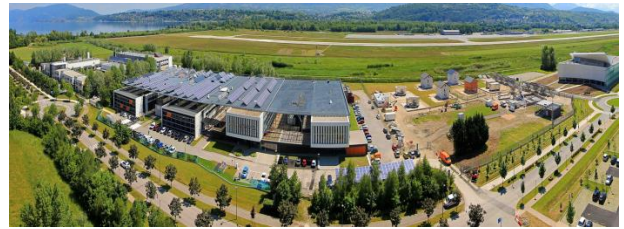

Figure 7: PRISMES platform.

Using the proposed architecture, a hybrid cloud server is envisaged in the De-Militarized Zone of PREDIS as the support for cross-infrastructure experiments. CIM is deployed over the cloud server using an adaptive approach. The platforms may choose to keep their own configuration and replicate the data to the cloud server in CIM format or to implement CIM in their infrastructure.

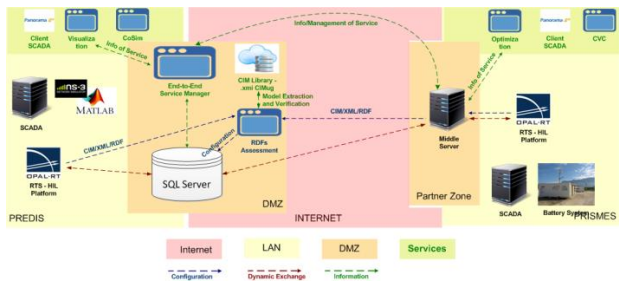

Figure 8: Interoperability architecture of PREDIS-PRISMES

## V. CONCLUSION AND OUTLOOK

In this paper, an adequate approach to assess holistic experiment of CPES is presented. By connecting existing and established infrastructures with complementary specialization and facilities into a cross-infrastructure holistic experiment over hybrid cloud based architecture, the proposition enables testing of CPES assessment research in near real-world scenario. The design of cross-infrastructure experiment is covered and an exemplary case-study of interoperability PREDIS-PRISMES is considered.

This architecture provides various experimental solutions: multi-platform co-simulation or remote Power-Hardware-in-the-loop, etc. These complement the required multi-domain complexity of validation framework for CPES assessment.

## ACKNOWLEDGEMENT

This research is sponsored by the French Carnot Institute "Energies du Futur" (http://www.energiesdufutur.eu/), in the PPInterop 2 project.

The participation from G2Elab and CEA INES is also partly supported by the European Community's Horizon 2020 Program (H2020/2014-2020) under project "ERIGrid" (Grant Agreement No. 654113, www.erigrid.eu).